\documentclass[twocolumn,showpacs,preprintnumbers,amsmath,amssymb]{revtex4}
\usepackage{graphicx}% Include figure files
\usepackage{dcolumn}% Align table columns on decimal point
\usepackage{bm}% bold math

\usepackage{amssymb}
\usepackage{amsmath}

\begin{document}

% \preprint{For PRD(\textbf{Ver.06})}

\title{Cosmological model based on both holographic-like connection and Padmanabhan's holographic equipartition law}
%%%\subtitle{}

\author{Nobuyoshi {\sc Komatsu}}  \altaffiliation{E-mail: komatsu@se.kanazawa-u.ac.jp} 
\affiliation{Department of Mechanical Systems Engineering, Kanazawa University, Kakuma-machi, Kanazawa, Ishikawa 920-1192, Japan}

%%\date{\today}

\begin{abstract}
A cosmological model based on holographic scenarios is formulated in a flat Friedmann--Robertson--Walker universe.
To formulate this model, the cosmological horizon is assumed to have a general entropy and a general temperature (including Bekenstein--Hawking entropy and Gibbons--Hawking temperature, respectively).
In addition, a holographic-like connection [Eur. Phys. J. C \textbf{83}, 690 (2023)] and Padmanabhan's holographic equipartition law are assumed for the entropy and temperature, and the Friedmann and acceleration equations are derived from these.
The derived Friedmann and acceleration equations include both the entropy and the temperature and are slightly complicated,
but can be combined into a single simple equation, corresponding to a similar equation that describes the background evolution of the universe in time-varying $\Lambda (t)$ cosmologies.
The simple equation depends on the entropy but not on the temperature because the temperatures in the Friedmann and acceleration equations cancel each other.
These results imply that the holographic-like connection should be consistent with Padmanabhan's holographic equipartition law through the present model and that the entropy plays a more important role.
When the Gibbons--Hawking temperature is used as the temperature, the Friedmann and acceleration equations are found to be equivalent to those for a $\Lambda(t)$ model.
A particular case of the present model is also examined, applying a power-law corrected entropy.

%%\keywords{First keyword \and Second keyword \and More}
%%\PACS{98.80.-k, 98.80.Es, 95.30.Tg}
\end{abstract}


\maketitle

\section{Introduction} 
\label{Introduction}

Many cosmological observations have implied an accelerated expansion of the late Universe \cite{PERL1998_Riess1998,Planck2018,Hubble2017}.
To explain the accelerated expansion, astrophysicists have proposed various cosmological models \cite{Bamba1Nojiri1Frusciante}, e.g., lambda cold dark matter ($\Lambda$CDM) models, time-varying $\Lambda (t)$ cosmology \cite{FreeseOverduin,Nojiri2006etc,Valent2015Sola2019,Sola_2009-2022}, bulk viscous models \cite{Weinberg0,BarrowLima,BrevikNojiri,EPJC2022}, creation of CDM (CCDM) models \cite{Prigogine_1988-1989,Lima1992-1996,Freaza2002Cardenas2020,LimaOthers2023}, 
and thermodynamic cosmological scenarios \cite{EassonCai,Basilakos1,Cai2005,Cai2008,Sheykhi1,Sheykhi2Karami,Koma456,Koma789,Neto2016,Neto2022}.
The famous $\Lambda$CDM model assumes an additional energy component called `dark energy'  in a Friedmann--Robertson--Walker (FRW) universe
or, equivalently, a cosmological constant $\Lambda$, corresponding to an extra driving term, added to the Friedmann and acceleration equations.

In contrast, thermodynamic cosmological scenarios are generally based on the `holographic principle', in which information for the bulk is assumed to be stored on the horizon \cite{Hooft-Bousso}.
(Here it is considered that the concept of black hole thermodynamics \cite{Hawking1Bekenstein1,GibbonsHawking1977} is applied to the cosmological horizon \cite{Easther1-Egan1,Pavon2013Mimoso2013,Bamba2018Pavon2019,deSitter_entropy,Saridakis20192021,Sheykhia2018,Jacob1995,Padma2010,Verlinde1,HDE,Padmanabhan2004,ShuGong2011,Koma14,Koma15,Koma16,Koma17,Koma19}.)
In those scenarios, Padmanabhan's holographic equipartition law \cite{Padma2012AB} has been examined from various viewpoints \cite{Cai2012,Moradpour,Hashemi,Wang,Koma10,Koma11,Koma12,Koma18,Krishna20172019,Mathew2022,Chen2022,Luciano,Mathew2023}.
Based on this law, the acceleration equation can be derived from the expansion of cosmic space due to the difference between the degrees-of-freedom on the surface and in the bulk in a region of space \cite{Padma2012AB}.

In addition, a holographic-like connection has been recently examined in standard cosmology \cite{Koma18}.
The holographic-like connection indicates that the Helmholtz free energy on the cosmological horizon is equivalent to the effective energy based on the energy density calculated from the Friedmann equation.
The connection is considered to be a bridge between horizon thermodynamics and cosmological equations \cite{Koma18}.
(Padmanabhan has derived a similar energy-balance relation, which is essentially equivalent to the holographic-like connection \cite{Pad2017}.)
By using the holographic-like connection, the Friedmann equation should be able to be derived from the free energy on the horizon.
Of course, when the continuity equation is assumed, the Friedmann equation can be derived from the acceleration equation and the continuity equation because two of the three equations are independent \cite{Ryden1}.
Also, the Friedmann equation can be derived from thermodynamics such as the first law of thermodynamics, see, e.g., Refs.\ \cite{Cai2005,Cai2008,Sheykhi1,Sheykhi2Karami} and recent works \cite{Sheykhia2018,ApparentHorizon2022}.
Moreover, the acceleration equation can be derived by applying the first law of thermodynamics to an apparent horizon, see, e.g., Ref.\ \cite{Cai2008}.
The holographic-like connection is thus considered to be a derivation method based on holographic scenarios.

In standard cosmology, the holographic-like connection is consistent with Padmanabhan's holographic equipartition law \cite{Koma18}.
Similarly, in modified FRW cosmologies, the holographic-like connection and the holographic equipartition law are expected to be consistent with each other. 
That is, the Friedmann equation from the connection should be consistent with the acceleration equation from the law, even in the modified FRW cosmology.
We expect that such a consistent cosmological model can be formulated using the Friedmann and acceleration equations based on the holographic scenario, namely the holographic-like connection and the holographic equipartition law.
However, from this viewpoint, cosmological models have not yet been extensively examined, though several works have been reported \cite{Tu2018,Tu2019}.
This model should provide new insights into discussions of holographic cosmology and horizon thermodynamics.

In this context, we phenomenologically formulate a cosmological model based on a holographic-like connection and Padmanabhan's holographic equipartition law.
In the present study, we do not select a particular entropy and a particular temperature, such as the Bekenstein--Hawking entropy \cite{Hawking1Bekenstein1} and the Gibbons--Hawking temperature \cite{GibbonsHawking1977}, because various forms of entropy \cite{Das2008,Radicella2010,LQG2004_123,Tsallis2012,Czinner1Czinner2,Barrow2020,Nojiri2022} and temperature \cite{Dynamical-T-20072014,Dynamical-T-1998,Dynamical-T-2008} have been proposed.
Instead, we consider a general entropy and a general temperature on the cosmological horizon.
Using the general entropy and temperature, we formulate a cosmological model based on the holographic scenario.
Accordingly, various forms of the entropy and temperature should be applied to the formulated model.
(This paper focusses on background evolution of the universe.)

The remainder of the present article is organized as follows.
In Sec.\ \ref{Horizon thermodynamics}, horizon thermodynamics is described.
In Sec.\ \ref{Entropy and temperature on the horizon}, the Bekenstein--Hawking entropy and Gibbons--Hawking temperature are reviewed and a general entropy and general temperature are introduced.
In Sec.\ \ref{Energy and free energy on the horizon}, the energy and Helmholtz free energy on the horizon are reviewed.
In Sec.\ \ref{Cosmological equations}, cosmological equations are reviewed and a holographic-like connection is introduced.
In Sec.\ \ref{Modified cosmological equations and the present model}, a cosmological model based on holographic scenarios is formulated.
In Sec.\ \ref{Modified Friedmann equation from a holographic-like connection}, a modified Friedmann equation is derived from the holographic-like connection.
In Sec.\ \ref{Padmanabhan's holographic equipartition law}, a modified acceleration equation is derived from Padmanabhan's holographic equipartition law.
Based on the Friedmann and acceleration equations, a cosmological model is formulated in Sec.\ \ref{The present model}.
In Sec.\ \ref{The present model with a power-law corrected entropy}, a particular case of the present model is examined, applying a power-law corrected entropy.
Finally, in Sec.\ \ref{Conclusions}, the conclusions of the study are presented.

The holographic-like connection has not yet been established, as discussed later.
However, the holographic-like connection is considered to be a viable scenario and, therefore, detailed studies are needed from various viewpoints. 
It should be worthwhile to examine cosmological models based on the holographic-like connection.

\section{Horizon thermodynamics} 
\label{Horizon thermodynamics}

The horizon thermodynamics is closely related to the holographic principle \cite{Hooft-Bousso}, which assumes that the horizon of the universe has an associated entropy and an approximate temperature \cite{EassonCai}.
The entropy and temperature are introduced in Sec.\ \ref{Entropy and temperature on the horizon} and
the energy and the Helmholtz free energy on the horizon are reviewed in Sec.\ \ref{Energy and free energy on the horizon}. 

For generality, a general entropy $S_{H}$ and general temperature $T_{H}$ on the horizon are considered so that various forms of the entropy and temperature can be applied to a formulated cosmological model.
In this study, we assume a homogeneous, isotropic, and spatially flat universe, namely a flat FRW universe.
Accordingly, the Hubble horizon is equivalent to an apparent horizon.
An expanding universe is also assumed.

\subsection{Entropy $S_{H}$ and temperature $T_{H}$ on the horizon} 
\label{Entropy and temperature on the horizon}

In this subsection, an entropy and a temperature on the Hubble horizon are introduced, according to previous works \cite{Jacob1995,Padma2010,Verlinde1,HDE,Padma2012AB,Cai2012,Moradpour,Hashemi,Wang,Sheykhia2018,Padmanabhan2004,ShuGong2011,Koma14,Koma15,Koma16,Koma17,Koma19}.

First, we review a form of the Bekenstein--Hawking entropy as an associated entropy on the cosmological horizon because it is the most standard.
In general, the cosmological horizon is examined by replacing the event horizon of a black hole by the cosmological horizon \cite{Koma17}. 
This replacement method has been widely accepted and we use it in this paper.
Based on the form of the Bekenstein--Hawking entropy, the entropy $S_{\rm{BH}}$ is written as \cite{Hawking1Bekenstein1}  
\begin{equation}
S_{\rm{BH}}  = \frac{ k_{B} c^3 }{  \hbar G }  \frac{A_{H}}{4}   ,
\label{eq:SBH}
\end{equation}
where $k_{B}$, $c$, $G$, and $\hbar$ are the Boltzmann constant, speed of light, gravitational constant, and reduced Planck constant, respectively.
The reduced Planck constant is defined by $\hbar \equiv h/(2 \pi)$, where $h$ is the Planck constant \cite{Koma14,Koma15,Koma16,Koma17,Koma19,Koma11,Koma12,Koma18}.
$A_{H}$ is the surface area of the sphere with Hubble horizon (radius) $r_{H}$ given by
\begin{equation}
     r_{H} = \frac{c}{H}   , 
\label{eq:rH}
\end{equation}
where the Hubble parameter $H$ is defined by 
\begin{equation}
   H \equiv   \frac{ da/dt }{a(t)} =   \frac{ \dot{a}(t) } {a(t)}  , 
\label{eq:Hubble}
\end{equation}
and $a(t)$ is the scale factor at time $t$.
Substituting $A_{H}=4 \pi r_{H}^2 $ into Eq.\ (\ref{eq:SBH}) and applying Eq.\ (\ref{eq:rH}) yields \cite{Koma14,Koma15,Koma16,Koma17,Koma19}
\begin{equation}
S_{\rm{BH}}  = \frac{ k_{B} c^3 }{  \hbar G }   \frac{A_{H}}{4}       
                  =  \left ( \frac{ \pi k_{B} c^5 }{ \hbar G } \right )  \frac{1}{H^2}  
                  =    \frac{K}{H^2}    , 
\label{eq:SBH2}      
\end{equation}
where $K$ is a positive constant given by
\begin{equation}
  K =  \frac{  \pi  k_{B}  c^5 }{ \hbar G } = \frac{  \pi  k_{B}  c^2 }{ L_{P}^{2} }   , 
\label{eq:K-def}
\end{equation}
and $L_{P}$ is the Planck length, written as
\begin{equation}
  L_{P} = \sqrt{ \frac{\hbar G} { c^{3} } }      .
\label{eq:Lp}
\end{equation}

Next, we review the Gibbons--Hawking temperature as an approximate temperature on the horizon.
The Gibbons--Hawking temperature $T_{\rm{GH}}$ is given by \cite{GibbonsHawking1977} 
\begin{equation}
T_{\rm{GH}}  = \frac{ \hbar H}{   2 \pi  k_{B}  }   .
\label{eq:T_H1}
\end{equation}
The above equation indicates that $T_{\rm{GH}}$ is proportional to $H$ and is constant during the evolution of de Sitter universes \cite{Koma17,Koma19}.
In this sense, the horizon of the de Sitter universe is considered to be static.

Equations\ (\ref{eq:SBH2}) and (\ref{eq:T_H1}) are widely used for the entropy and temperature on the horizon, respectively \cite{Jacob1995,Padma2010,Verlinde1,HDE,Padma2012AB,Cai2012,Moradpour,Hashemi,Wang,Sheykhia2018,Padmanabhan2004,ShuGong2011,Koma14,Koma15,Koma16,Koma17,Koma19}.
In this study, for generality, we consider a general entropy $S_{H}$ and general temperature $T_{H}$, written as 
\begin{equation}
  S_{H} = S_{\rm{BH}}  \left ( \frac{ S_{H} }{ S_{\rm{BH}} }  \right )    \quad \textrm{and} \quad      T_{H} = T_{\rm{GH}} \left ( \frac{T_{H}}{T_{\rm{GH}}} \right )   , 
\label{eq:S_H_T_H}
\end{equation}
where $S_{H}$ and $T_{H}$ include $(S_{H} / S_{\rm{BH}})$ and $(T_{H} / T_{\rm{GH}})$, respectively.
We use the above notation because it is suitable for formulating a cosmological model, as examined later.

\subsection{Energy $E_{H}$ and free energy $F_{H}$ on the horizon} 
\label{Energy and free energy on the horizon}

In this subsection, we review the energy $E_{H}$ and Helmholtz free energy $F_{H}$ on the horizon, based on a previous work \cite{Koma18}.
Hereafter, we refer to $F_{H}$ as the free energy (on the horizon).

We have assumed that information for the bulk is stored on the horizon based on the holographic principle.  
We now assume the equipartition law of energy on the horizon, according to Refs.\ \cite{Padma2010,ShuGong2011}.
Consequently, an energy on the Hubble horizon, $E_{H}$, can be written as
\begin{equation}
E_{H} =  N_{\rm{sur}}  \times \frac{1}{2} k_{B} T_{H}     , 
\label{E_equip}
\end{equation}
where $N_{\rm{sur}}$ is the number of degrees of freedom on a spherical surface of Hubble radius $r_{H}$ and is written as \cite{Koma14}
\begin{equation}
  N_{\rm{sur}} = \frac{4 S_{H} }{k_{B}}       .
\label{N_sur}
\end{equation}
Substituting Eq.\ (\ref{N_sur}) into Eq.\ (\ref{E_equip}) yields
\begin{equation}
E_{H} =  \left ( \frac{4 S_{H} }{k_{B}} \right )     \frac{1}{2} k_{B} T_{H}  =  2 S_{H}  T_{H}  . 
\label{E_ST2_thermo_2}
\end{equation}
This thermodynamic relation, namely ${E}_{H}   =2 S_{H}  T_{H} $, was proposed by Padmanabhan \cite{Padmanabhan2004,Padma2010}.
(In Ref.\ \cite{Koma18}, the same relation was discussed, using $dE_{H} =T_{H} dS_{H}$.)

Based on thermodynamics, the free energy $F_{H}$ on the horizon can be defined as
\begin{equation}
   F_{H} = E_{H} - T_{H} S_{H}  .
\label{F_def_thermo}
\end{equation}
Substituting $T_{H} S_{H} = E_{H}/2$ given by Eq.\ (\ref{E_ST2_thermo_2}) into Eq.\ (\ref{F_def_thermo}) yields 
\begin{equation}
   F_{H} = E_{H} - T_{H} S_{H} = E_{H} - \frac{1}{2}E_{H} = \frac{1}{2}E_{H}  .
\label{F_2_thermo}
\end{equation}
The free energy $F_{H}$ is half of $E_{H}$ \cite{Koma18}.

When both $S_{H} = S_{\rm{BH}}$ and $T_{H} = T_{\rm{GH}}$ are considered, the free energy $F_{H}$ is written as \cite{Koma18}
\begin{equation}
   F_{H}  = \frac{1}{2}E_{H}  = S_{H}  T_{H}  = S_{\rm{BH}} T_{\rm{GH}} =  \frac{1}{2} \frac{ c^{5} }{ G }  \left ( \frac{1}{H} \right ) ,
\label{F_3_thermo}
\end{equation}
where Eqs.\ (\ref{eq:SBH2}) and (\ref{eq:T_H1}) have been used. 
The obtained free energy is related to an effective energy in the Hubble volume \cite{Koma18}.
This relation, namely a holographic-like connection, has not yet been established and is discussed in the next section.

\section{Cosmological equations and a holographic-like connection} 
\label{Cosmological equations}

We consider a flat FRW universe.
In Sec.\ \ref{A general formulation}, cosmological equations for a $\Lambda(t)$ model similar to time-varying $\Lambda(t)$ cosmologies are reviewed.
In Sec.\ \ref{A holographic-like connection in standard cosmology}, a holographic-like connection in standard cosmology is introduced.

\subsection{General formulation for a $\Lambda(t)$ model} 
\label{A general formulation}

A cosmological model based on holographic scenarios examined in the present study is expected to be related to a $\Lambda(t)$ model.
(This is discussed in Sec.\ \ref{The present model}.)
Therefore, in this subsection, we introduce a general formulation for cosmological equations for the $\Lambda(t)$ model, according to previous works \cite{Koma14,Koma15,Koma16}.
The general Friedmann and acceleration equations are written as 
\begin{equation}
 H(t)^2      =  \frac{ 8\pi G }{ 3 } \rho (t)    + f_{\Lambda}(t)            ,                                                 
\label{eq:General_FRW01_f_0} 
\end{equation} 
\begin{align}
  \frac{ \ddot{a}(t) }{ a(t) }     &= \dot{H} + H^{2} = -  \frac{ 4\pi G }{ 3 }  ( 1+  3w ) \rho (t)            +   f_{\Lambda}(t)      ,  
\label{eq:General_FRW02_f_0}
\end{align}
where $\rho(t)$ and $p(t)$ are the mass density of cosmological fluids and the pressure of cosmological fluids, respectively. 
$w$ represents the equation-of-state parameter for a generic component of matter, which is given as \cite{Koma14,Koma15,Koma16}
\begin{equation}
  w = \frac{ p(t) } { \rho(t)  c^2 }    .
\label{eq:w}
\end{equation}
For a $\Lambda$-dominated universe, matter-dominated universe, and radiation-dominated universe, $w$ is $-1$, $0$, and $1/3$, respectively.
An extra driving term, $f_{\Lambda}(t)$, is phenomenologically assumed.
(The continuity equation is discussed in Appendix\ \ref{Continuity equation}.)

From the Friedmann and acceleration equations, we can obtain a simple equation that describes the background evolution of the universe.
Combining Eq.\ (\ref{eq:General_FRW01_f_0}) with Eq.\ (\ref{eq:General_FRW02_f_0}) yields \cite{Koma14}
\begin{align}
      \dot{H} &= - \frac{3}{2} (1+w) H^{2}  +  \frac{3}{2} (1+w)    f_{\Lambda}(t)                  \notag \\
                 &= - \frac{3}{2} (1+w) H^{2}  \left ( 1-   \frac{ f_{\Lambda}(t) }{H^{2}} \right )           .
\label{eq:Back_f}
\end{align}
By solving this equation, we can examine the background evolution of the universe for a $\Lambda(t)$ model.
In Secs.\ \ref{Modified cosmological equations and the present model} and \ref{The present model with a power-law corrected entropy}, we compare Eq.\ (\ref{eq:Back_f}) and an equation obtained from a cosmological model based on holographic scenarios.

\subsection{A holographic-like connection}
\label{A holographic-like connection in standard cosmology}

In this subsection, we introduce a holographic-like connection in standard cosmology, according to Ref.\ \cite{Koma18}.
For standard cosmology, we set $f_{\Lambda}(t) =0$.
Consequently, Eqs.\ (\ref{eq:General_FRW01_f_0}) and (\ref{eq:General_FRW02_f_0}) can be written as
\begin{equation}
 H^2      =  \frac{ 8\pi G }{ 3 } \rho              ,                                                 
\label{eq:FRW01} 
\end{equation} 
\begin{align}
  \frac{ \ddot{a} }{ a }  =  \dot{H} + H^{2}   = -  \frac{ 4\pi G }{ 3 }  ( 1+  3w ) \rho            .
\label{eq:FRW02}
\end{align}

We now introduce a holographic-like connection between thermostatistical quantities on a cosmological horizon and in the bulk \cite{Koma18}.
To this end, we determine the effective energy based on the energy density calculated from the Friedmann equation.
Using the Friedmann equation given by Eq.\ (\ref{eq:FRW01}), the energy density $\rho c^{2}$ is written as \cite{Koma18}
\begin{equation}
 \rho c^{2}  =    \frac{ 3 c^{2} }{ 8 \pi G }  H^{2}   .
\label{Energy-density_FRW}
\end{equation}
The energy density $\rho c^{2}$ is generally defined by 
\begin{equation}
 \rho c^{2}  =    \frac{ E_{\rm{eff}} }{ V }, 
\label{Energy-density_EeffV}
\end{equation}
where $E_{\rm{eff}}$ is the effective energy and $V$ is the volume of a sphere with the Hubble horizon (radius) \cite{Koma18},
given by
\begin{equation}
V = \frac{4 \pi}{3} r_{H}^{3} =  \frac{4 \pi}{3} \left ( \frac{c}{H} \right )^{3}   .
\label{eq:V}
\end{equation}
Solving Eq.\ (\ref{Energy-density_EeffV}) with respect to $E_{\rm{eff}}$ and substituting Eqs.\ (\ref{Energy-density_FRW}) and (\ref{eq:V}) into the resultant equation yields \cite{Koma18}
\begin{align}
E_{\rm{eff}}  &= \rho c^{2} V =  \frac{ 3 c^{2} }{ 8 \pi G }  H^{2}    \frac{4}{3} \pi \left ( \frac{c}{H} \right )^3   = \frac{1}{2}   \frac{ c^{5} }{ G }  \left ( \frac{1}{H} \right )       .
\label{Eeff_FRW}
\end{align}
The above equation is equivalent to Eq.\ (\ref{F_3_thermo}).
Therefore, the free energy $F_{H}$ on the Hubble horizon is equivalent to the effective energy $E_{\rm{eff}}$ in the Hubble volume \cite{Koma18}:
\begin{equation}
  F_{H}  = E_{\rm{eff}}       ,
\label{F_Eeff_thermo_2}
\end{equation}
where $S_{H} = S_{\rm{BH}}$ and $T_{H} = T_{\rm{GH}}$ are considered.
This consistency is a `holographic-like connection' in standard cosmology.
The holographic-like connection, namely $F_{H} = E_{\rm{eff}}$, is expected to be a bridge between horizon thermodynamics and cosmological equations \cite{Koma18}.
A similar energy-balance relation $\rho c^2 V = T_{\rm{GH}} S_{\rm{BH}}$ was derived by Padmanabhan \cite{Pad2017} and was described in, e.g., Refs.\ \cite{Tu2018,Tu2019}.
The energy balance relation is essentially equivalent to the holographic-like connection.
However, the free energy and the holographic-like connection $F_{H} = E_{\rm{eff}}$ were not discussed in those works.

The holographic-like connection is considered to be a viable scenario.
Of course, in the above discussion, a standard FRW cosmology is assumed and, in addition, the Bekenstein--Hawking entropy and the Gibbons--Hawking temperature are selected as an associated entropy and an approximate temperature on the horizon, respectively.
In fact, the selected forms of the entropy and temperature are expected to lead to modified FRW cosmologies, as examined in Ref.\ \cite{ApparentHorizon2022}.
That is, extended forms of the entropy and temperature should be suitable for discussions of the holographic-like connection in the modified FRW cosmology.
Therefore, we consider a general entropy $S_{H}$ and a general temperature $T_{H}$ and assume that the holographic-like connection can be applied to $S_{H}$ and $T_{H}$.
In the next section, we derive a modified Friedmann equation from the holographic-like connection and formulate a cosmological model.

Before proceeding further, we explain the holographic-like connection again.
The holographic-like connection implies that the free energy on the horizon (which is obtained from the equipartition law of energy) is considered to be equal to the effective energy, namely the bulk energy \cite{Koma18}.
However, the holographic-like connection has not yet been established.
In fact, the energy in the equipartition law has been discussed from different viewpoints.
For example, Verlinde argued that the energy in the equipartition law should be related to the total enclosed mass, to derive Newton's law of gravitation \cite{Verlinde1}.
Also, Padmanabhan argued that the equipartition energy on the horizon should be related to active gravitational mass \cite{Padma2010} and that the bulk energy should be the Komar energy \cite{Padma2012AB}.
On the other hand, Padmanabhan derived an energy-balance relation \cite{Pad2017}, essentially equivalent to the holographic-like connection.
Accordingly, the holographic-like connection may be related to a realistic holographic connection.
Detailed studies are needed from various viewpoints.
In the present paper, we accept the holographic-like connection and examine cosmological models based on the connection.

\section{Modified cosmological equations and the present model}
\label{Modified cosmological equations and the present model}

In this section, we consider a general entropy $S_{H}$ and general temperature $T_{H}$ and formulate a cosmological model based on both a holographic-like connection and Padmanabhan's holographic equipartition law.
In Sec.\ \ref{Modified Friedmann equation from a holographic-like connection}, a modified Friedmann equation is derived from the holographic-like connection.
In Sec.\ \ref{Padmanabhan's holographic equipartition law}, a modified acceleration equation is derived from the holographic equipartition law.
In Sec.\ \ref{The present model}, a cosmological model is formulated based on the derived Friedmann and acceleration equations.
(A modified continuity equation is discussed in Appendix\ \ref{Continuity equation}.)

The holographic-like connection, Padmanabhan's holographic equipartition law, and several assumptions used for the present model have not yet been established but are considered to be viable scenarios,
and accepted as such in this work.
(For example, we assume that the connection and the law can be applied to $S_{H}$ and $T_{H}$.)

\subsection{Modified Friedmann equation from a holographic-like connection}
\label{Modified Friedmann equation from a holographic-like connection}

In this subsection, we derive a modified Friedmann equation from a holographic-like connection.
From Eq.\ (\ref{F_Eeff_thermo_2}), the holographic-like connection is written as  
\begin{equation}
  F_{H}  = E_{\rm{eff}}     .
\label{F_Eeff_thermo_3}
\end{equation}
Originally, $S_{\rm{BH}}$ and $T_{\rm{GH}}$ are considered.
In the present study, the holographic-like connection $F_{H} = E_{\rm{eff}}$ is extended as if it is a basic principle.
That is, we assume that the holographic-like connection can be applied to a general entropy $S_{H}$ and general temperature $T_{H}$.
Based on this assumption, we derive the Friedmann equation from $F_{H} = E_{\rm{eff}}$ given by Eq.\ (\ref{F_Eeff_thermo_3}).

We first calculate the left-hand side of Eq.\ (\ref{F_Eeff_thermo_3}), namely, $F_{H}$.
From Eqs.\ (\ref{E_ST2_thermo_2}) and (\ref{F_2_thermo}), the free energy $F_{H}$ on the horizon is given by 
\begin{equation}
   F_{H} = \frac{1}{2}E_{H} = S_{H}  T_{H}  .
\label{F_thermo_3}
\end{equation}
Substituting Eq.\ (\ref{eq:S_H_T_H}) into this equation yields
\begin{equation}
   F_{H} =  S_{H}  T_{H}  = S_{\rm{BH}}  \left ( \frac{ S_{H} }{ S_{\rm{BH}} }  \right ) T_{\rm{GH}} \left ( \frac{T_{H}}{T_{\rm{GH}}} \right ) .
\label{F_SHTH}
\end{equation}
In addition, substituting Eqs.\ (\ref{eq:SBH2}) and (\ref{eq:T_H1}) into Eq.\ (\ref{F_SHTH}) yields
\begin{align}
   F_{H} &= S_{\rm{BH}}  \left ( \frac{ S_{H} }{ S_{\rm{BH}} }  \right ) T_{\rm{GH}} \left ( \frac{T_{H}}{T_{\rm{GH}}} \right )  \notag \\
           &=  \left ( \frac{ \pi k_{B} c^5 }{ \hbar G } \right )  \frac{1}{H^2}  \left ( \frac{ S_{H} }{ S_{\rm{BH}} }  \right )  \frac{ \hbar H}{   2 \pi  k_{B}  }  \left ( \frac{T_{H}}{T_{\rm{GH}}} \right )  \notag \\
           &=  \left ( \frac{ c^5 }{ 2 G }  \frac{1}{H} \right )  \left ( \frac{ S_{H} }{ S_{\rm{BH}} }  \right )   \left ( \frac{T_{H}}{T_{\rm{GH}}} \right )  .
\label{F_SHTH_2}
\end{align}
The factors $(S_{H} / S_{\rm{BH}})$ and $(T_{H} / T_{\rm{GH}})$ are retained, to allow a comparison between a derived Friedmann equation and the Friedmann equation in standard cosmology.

We next calculate the right-hand side of Eq.\ (\ref{F_Eeff_thermo_3}), namely $E_{\rm{eff}}$.
For this, we use $E_{\rm{eff}} = \rho c^{2} V$ given by Eq.\ (\ref{Energy-density_EeffV}).
From Eqs.\ (\ref{Energy-density_EeffV}) and (\ref{eq:V}), $E_{\rm{eff}}$ is given by 
\begin{align}
E_{\rm{eff}} &= \rho c^{2} V  = \rho c^{2} \frac{4 \pi}{3} r_{H}^{3} = \rho c^{2} \frac{4 \pi}{3} \left ( \frac{c}{H} \right )^{3} .
\label{Eeff_1}
\end{align}

We now calculate Eq.\ (\ref{F_Eeff_thermo_3}). 
Substituting Eqs.\ (\ref{F_SHTH_2}) and (\ref{Eeff_1}) into Eq.\ (\ref{F_Eeff_thermo_3}) yields
\begin{align}
 \left ( \frac{ c^5 }{ 2 G }  \frac{1}{H} \right )  \left ( \frac{ S_{H} }{ S_{\rm{BH}} }  \right )   \left ( \frac{T_{H}}{T_{\rm{GH}}} \right )  &=  \rho c^{2} \frac{4 \pi}{3} \left ( \frac{c}{H} \right )^{3} .
\label{F_Eeff_1}
\end{align}
The above equation can be written as
\begin{align}
   H^{2} \left ( \frac{ S_{H} }{ S_{\rm{BH}} }  \right )   \left ( \frac{T_{H}}{T_{\rm{GH}}} \right )  &=  \frac{8 \pi G}{3} \rho   ,
\label{ModifiedFriedmann_1}
\end{align}
or equivalently
\begin{align}
  H^{2}    &=  \frac{8 \pi G}{3} \rho   + H^{2}   \left [ 1-  \left ( \frac{ S_{H} }{ S_{\rm{BH}} }  \right )   \left ( \frac{T_{H}}{T_{\rm{GH}}} \right )  \right ]  .
\label{ModifiedFriedmann_1_equiv}
\end{align}
These are the modified Friedmann equation derived from the holographic-like connection.
Equations\ (\ref{ModifiedFriedmann_1}) and (\ref{ModifiedFriedmann_1_equiv}) indicate that both the normalized entropy $S_{H} / S_{\rm{BH}}$ and normalized temperature $T_{H} / T_{\rm{GH}}$ affect the Friedmann equation.
When both $S_{H} = S_{\rm{BH}}$ and $T_{H} = T_{\rm{GH}}$ are considered, Eqs.\ (\ref{ModifiedFriedmann_1}) and (\ref{ModifiedFriedmann_1_equiv}) reduce to Eq.\ (\ref{eq:FRW01}), namely the Friedmann equation in standard cosmology.
Equation\ (\ref{ModifiedFriedmann_1_equiv}) includes the second term on the right-hand side, corresponding to an extra driving term.
We can interpret that this driving term is implicitly included in Eq.\ (\ref{ModifiedFriedmann_1}), because Eq.\ (\ref{ModifiedFriedmann_1}) is equivalent to Eq.\ (\ref{ModifiedFriedmann_1_equiv}).
In the present study, Eq.\ (\ref{ModifiedFriedmann_1_equiv}) is typically used, instead of Eq.\ (\ref{ModifiedFriedmann_1}).

In this subsection, we have derived the modified Friedmann equation from the holographic-like connection.
In the next subsection, a modified acceleration equation is derived from Padmanabhan's holographic equipartition law.

\subsection{Modified acceleration equation from Padmanabhan's holographic equipartition law} 
\label{Padmanabhan's holographic equipartition law}

Based on Padmanabhan's holographic equipartition law, the acceleration equation can be derived from the expansion of cosmic space due to the difference between the degrees-of-freedom (DOF) on the surface and in the bulk in a region of space \cite{Padma2012AB}.
The law is reviewed in previous works \cite{Cai2012,Sheykhia2018,Koma10,Koma11,Koma12,Koma18}.
In this subsection, based on those works, we introduce Padmanabhan's holographic equipartition law and derive a modified acceleration equation from it.

In an infinitesimal interval $dt$ of cosmic time, the increase $dV$ in the Hubble volume can be expressed as \cite{Padma2012AB}
\begin{equation}
     \frac{dV}{dt}  =  L_{p}^{2} (N_{\rm{sur}} - \epsilon N_{\rm{bulk}} ) \times c      , 
\label{dVdt_N-N}
\end{equation}
where $N_{\rm{sur}}$ is the number of DOF on a spherical surface of Hubble radius $r_{H}$ and $N_{\rm{bulk}}$ is the number of DOF in the bulk. 
$L_{p}$ is the Planck length given by Eq.\ (\ref{eq:Lp}) and $\epsilon$ is a parameter defined as \cite{Padma2012AB}
 \begin{equation}
        \epsilon \equiv     
 \begin{cases}
              +1  & (\rho c^2 + 3p <0  \textrm{: an accelerating universe}),  \\ 
              -1  & (\rho c^2 + 3p >0   \textrm{: a decelerating universe}).    \\
\end{cases}
\label{epsilon}
\end{equation}
We refer to Eq.\ (\ref{dVdt_N-N}) as Padmanabhan's holographic equipartition law.
From this equation, the acceleration equation can be derived based on previous reports \cite{Koma10,Koma11,Koma12,Koma18}.
In this study, a general temperature $T_{H}$ is considered, by extending the derivation in those reports.

To derive the acceleration equation, we first calculate the left-hand side of Eq.\ (\ref{dVdt_N-N}), namely $dV/dt$. 
Differentiating Eq.\ (\ref{eq:V}) with respect to $t$ yields \cite{Koma10,Koma11,Koma12,Koma18}
\begin{equation}
     \frac{dV}{dt}  =   \frac{d}{dt} \left ( \frac{4 \pi}{3} \left ( \frac{c}{H} \right )^{3}  \right )   =  -4 \pi c^{3}   \left (  \frac{ \dot{H} }{H^{4} } \right )  , 
\label{dVdt_right}
\end{equation}
where $r$ is set to $r_{H}=c/H$ before the time derivative is calculated \cite{Padma2012AB}.   
Next, to calculate the right-hand side of Eq.\ (\ref{dVdt_N-N}), the number of DOF in the bulk $N_{\rm{bulk}}$ is assumed to obey the equipartition law of energy \cite{Padma2012AB}: 
\begin{equation}
  N_{\rm{bulk}} = \frac{|E_{\rm{bulk}}|}{ \frac{1}{2} k_{B} T_{H}   }     ,
\label{N_bulk_00}
\end{equation}
where $T_{H}$ is the general temperature on the horizon.
The Komar energy $|E_{\rm{bulk}}|$ contained inside the Hubble volume $V$ is assumed to be given by \cite{Padma2012AB}
\begin{equation}
|E_{\rm{bulk}}| =  |( \rho c^2 + 3p)| V  = - \epsilon ( \rho c^2 + 3p) V  .
\label{Komar}
\end{equation}
In addition, Eq.\ (\ref{N_bulk_00}) can be written as
\begin{equation}
  N_{\rm{bulk}}  = \frac{|E_{\rm{bulk}}|}{ \frac{1}{2} k_{B} T_{H} }  = \frac{|E_{\rm{bulk}}|}{ \frac{1}{2} k_{B} T_{\rm{GH}} \left (  \frac{T_{H} }{T_{\rm{GH}}} \right ) }     ,
\label{N_bulk}
\end{equation}
where $T_{\rm{GH}}$ is the Gibbons--Hawking temperature given by Eq.\ (\ref{eq:T_H1}).
Also, from Eq.\ (\ref{N_sur}), the number of DOF on the spherical surface $N_{\rm{sur}}$ is given by  
\begin{equation}
  N_{\rm{sur}} = \frac{4 S_{H} }{k_{B}}       . 
\label{N_sur_P}
\end{equation}

We now derive a modified acceleration equation from the holographic equipartition law.
According to Ref.\ \cite{Padma2012AB}, $\rho c^2 + 3p <0$ is selected and, therefore, $\epsilon = +1$ from Eq.\ (\ref{epsilon}).
This selection does not affect the following result.
We first calculate $N_{\rm{bulk}}$ on the right-hand side of Eq.\ (\ref{dVdt_N-N}).
Substituting Eqs.\ (\ref{eq:T_H1}) and (\ref{Komar}) into Eq.\ (\ref{N_bulk}) and using Eqs.\ (\ref{eq:w}) and (\ref{eq:V}) and $\epsilon = +1$ yields \cite{Koma10,Koma11,Koma12,Koma18}
\begin{align}
  N_{\rm{bulk}}   &= \frac{|E_{\rm{bulk}}|}{ \frac{1}{2} k_{B} T_{\rm{GH}} \left (  \frac{T_{H} }{T_{\rm{GH}}} \right ) }  \notag \\
                       &=   -  \frac{ (4 \pi)^{2} c^{5}  }{ 3 \hbar }  (1+3w)\rho  \frac{1}{  H^{4}   } \left (  \frac{T_{\rm{GH}}}{T_{H}} \right )      ,
\label{N_bulk_cal}
\end{align}
where $(T_{\rm{GH}}/T_{H})$ is retained, to allow a comparison between a derived acceleration equation and the acceleration equation examined in Refs.\ \cite{Koma10,Koma11,Koma12,Koma18}.
Substituting $\epsilon = +1$ and Eqs.\ (\ref{eq:Lp}), (\ref{dVdt_right}), (\ref{N_sur_P}), and (\ref{N_bulk_cal}) into Eq.\ (\ref{dVdt_N-N}) and solving the resultant equation with respect to $\dot{H}$ yields \cite{Koma10,Koma11,Koma12,Koma18}  
\begin{align}
  \dot{H}                                                     
                  &=   -  \frac{ 4 \pi G }{ 3} (1+3w)  \rho \left (  \frac{T_{\rm{GH}}}{T_{H}} \right )        - \frac{ S_{H} H^{4} }{K}                        , 
\label{dVdt_N-N_cal2}
\end{align}
where $K$ is given by Eq.\ (\ref{eq:K-def}).
In addition, substituting Eq.\ (\ref{dVdt_N-N_cal2}) into $ \ddot{a}/ a   =  \dot{H} + H^{2}$ and using $S_{\rm{BH}} = K/H^{2}$ given by Eq.\ (\ref{eq:SBH2}) yields \cite{Koma10,Koma11,Koma12,Koma18}
\begin{align}
  \frac{ \ddot{a} }{ a }       &=  \dot{H} + H^{2}             \notag \\
                                      &=   -  \frac{ 4 \pi G }{ 3}  (1+3w)  \rho \left (  \frac{T_{\rm{GH}}}{T_{H}} \right )       - \frac{ S_{H} H^{4} }{K}      + H^{2}  \notag \\
                                      &=   -  \frac{ 4 \pi G }{ 3}  (1+3w)  \rho \left (  \frac{T_{\rm{GH}}}{T_{H}} \right )      + H^{2}  \left (  1 - \frac{ S_{H} }{ S_{\rm{BH}} }  \right )     .
\label{N-N_FRW02_SH}
\end{align}
This is the modified acceleration equation derived from Padmanabhan's holographic equipartition law. 
The first and second terms on the right-hand side include $T_{\rm{GH}}/T_{H}$ and $S_{H} / S_{\rm{BH}}$, respectively.
Except for $T_{\rm{GH}}/T_{H}$, Eq.\ (\ref{N-N_FRW02_SH}) is the same as the acceleration equation examined in Refs.\ \cite{Koma10,Koma11,Koma12,Koma18}.
We note that $T_{\rm{GH}}/T_{H}$ is written as $(T_{H}/T_{\rm{GH}})^{-1}$, using the normalized temperature $T_{H}/T_{\rm{GH}}$.

When $S_{H} \neq S_{\rm{BH}}$, the second term $H^{2}(1- S_{H}/S_{\rm{BH}})$ on the right-hand side of Eq.\ (\ref{N-N_FRW02_SH}) is non-zero.
In contrast, when $S_{H} = S_{\rm{BH}}$, the second term is zero.
In addition, when both $S_{H} = S_{\rm{BH}}$ and $T_{H} = T_{\rm{GH}}$, Eq.\ (\ref{N-N_FRW02_SH}) reduces to Eq.\ (\ref{eq:FRW02}), namely the acceleration equation in standard cosmology.

In this subsection, we have derived the modified acceleration equation from Padmanabhan's holographic equipartition law.
In the next subsection, we formulate a cosmological model using the derived Friedmann and acceleration equations.

\subsection{The present model}
\label{The present model}

We have derived the modified Friedmann and acceleration equations from the holographic-like connection and Padmanabhan's holographic equipartition law, respectively.
In this subsection, we phenomenologically formulate a cosmological model based on the derived Friedmann and acceleration equations.
For this, we assume that the holographic-like connection and the holographic equipartition law are consistent with each other, through the holographic scenario.
This consistency is discussed later.
(A similar model was examined in the works of Tu \textit{et al}. \cite{Tu2018,Tu2019}, using the Bekenstein--Hawking entropy.)

First, the Friedmann and acceleration equations for the present model are summarized.
(The continuity equation derived from the Friedmann and acceleration equations is discussed in Appendix\ \ref{Continuity equation}.)
From Eq.\ (\ref{ModifiedFriedmann_1_equiv}), the Friedmann equation is written as    
\begin{align}
  H^{2}    &=  \frac{8 \pi G}{3} \rho   + H^{2}   \left [ 1-  \left ( \frac{ S_{H} }{ S_{\rm{BH}} }  \right )   \left ( \frac{T_{H}}{T_{\rm{GH}}} \right )  \right ]  .
\label{Modified_FRW01_equiv_2}
\end{align}
From Eq.\ (\ref{N-N_FRW02_SH}), the acceleration equation is written as  
\begin{align}
  \frac{ \ddot{a} }{ a }   &=  \dot{H} + H^{2}             \notag \\
                                  &=   -  \frac{ 4 \pi G }{ 3}  (1+3w)  \rho \left (  \frac{T_{\rm{GH}}}{T_{H}} \right )      + H^{2}  \left (  1 - \frac{ S_{H} }{ S_{\rm{BH}} }  \right )     .
\label{Modified_FRW02_2}
\end{align}
When both $S_{H} = S_{\rm{BH}}$ and $T_{H} = T_{\rm{GH}}$ are considered, the Friedmann and acceleration equations reduce to Eqs.\ (\ref{eq:FRW01}) and (\ref{eq:FRW02}), respectively.
Of course, in general, the Friedmann and acceleration equations should be slightly complicated because of coefficients related to the entropy and the temperature.
However, we can obtain a simple equation from these two equations as follows.

Using Eq.\ (\ref{Modified_FRW01_equiv_2}), $\rho$ is given by
\begin{align}
    \rho  &=   \frac{3 H^{2} }{8 \pi G}  \left ( \frac{ S_{H} }{ S_{\rm{BH}} }  \right )   \left ( \frac{T_{H}}{T_{\rm{GH}}} \right )  .
\label{rho_2}
\end{align}
Substituting Eq.\ (\ref{rho_2}) into Eq.\ (\ref{Modified_FRW02_2}) yields
\begin{align}
   & \dot{H} + H^{2} =   -  \frac{ 4 \pi G }{ 3}  (1+3w)  \rho \left (  \frac{T_{\rm{GH}}}{T_{H}} \right )      + H^{2}  \left (  1 - \frac{ S_{H} }{ S_{\rm{BH}} }  \right )     \notag \\
   &=   -  \frac{ 4 \pi G }{ 3}  (1+3w)  \left [ \frac{3 H^{2} }{8 \pi G} \left ( \frac{ S_{H} }{ S_{\rm{BH}} }  \right )   \left ( \frac{T_{H}}{T_{\rm{GH}}} \right )  \right ]  \left (  \frac{T_{\rm{GH}}}{T_{H}} \right )   \notag \\
   &  \quad  + H^{2}  \left (  1 - \frac{ S_{H} }{ S_{\rm{BH}} }  \right )     \notag \\
   &=   -  \frac{1}{2}  (1+3w)  H^{2} \left ( \frac{ S_{H} }{ S_{\rm{BH}} }  \right )   + H^{2}  \left (  1 - \frac{ S_{H} }{ S_{\rm{BH}} }  \right )   \notag \\
   &=   -  \frac{3}{2}  (1+ w)  H^{2} \left ( \frac{ S_{H} }{ S_{\rm{BH}} }  \right )   + H^{2}   .
\label{Modified_eq_0}
\end{align}
The temperature is cancelled in this calculation.
From the above equation, a simple equation can be obtained: 
\begin{align}
    \dot{H}  =  - \frac{3}{2}  (1+ w)  H^{2} \left ( \frac{ S_{H} }{ S_{\rm{BH}} }  \right )  .
\label{Modified_eq_1}
\end{align}
This equation is derived from the Friedmann and acceleration equations for the present model.
By solving Eq.\ (\ref{Modified_eq_1}), we can examine the background evolution of the universe.
That is, Eq.\ (\ref{Modified_eq_1}) corresponds to Eq.\ (\ref{eq:Back_f}) for a $\Lambda(t)$ model.
These results imply that the holographic-like connection should be consistent with Padmanabhan's holographic equipartition law.
This consistency may be interpreted as a kind of holographic duality.

Interestingly, Eq.\ (\ref{Modified_eq_1}) depends on $S_{H}$ and does not depend on $T_{H}$, although the Friedmann and acceleration equations include both $S_{H}$ and $T_{H}$.
This is because $T_{H}$ included in these two equations cancel each other.
Equation\ (\ref{Modified_eq_1}) indicates that the background evolution of the universe for the present model depends on $S_{H}$ but does not depend on $T_{H}$.
Accordingly, the entropy probably plays a more important role in the present model.
Based on the present model, we can observe the evolution of a temperature which is different from the selected $T_{H}$, as examined in the next section.

 In this section, we consider two fundamental cases: $S_{H} = S_{\rm{BH}}$ and $T_{H}=T_{\rm{GH}}$,
where $S_{\rm{BH}}$ is the Bekenstein--Hawking entropy and $T_{\rm{GH}}$ is the Gibbons--Hawking temperature.

\subsubsection{$S_{H} = S_{\rm{BH}}$ (Bekenstein--Hawking entropy)}

When $S_{H} = S_{\rm{BH}}$, Eqs.\ (\ref{Modified_FRW01_equiv_2}) and (\ref{Modified_FRW02_2}) reduce to
\begin{align}
  H^{2}    &=  \frac{8 \pi G}{3} \rho   + H^{2}   \left ( 1-    \frac{T_{H}}{T_{\rm{GH}}}  \right )  ,
\label{Modified_FRW01_SH=SBH}
\end{align}
and
\begin{align}
  \frac{ \ddot{a} }{ a }       &=   -  \frac{ 4 \pi G }{ 3}  (1+3w)  \rho \left (  \frac{T_{\rm{GH}}}{T_{H}} \right )    .
\label{Modified_FRW02_SH=SBH}
\end{align}
Equations\ (\ref{Modified_FRW01_SH=SBH}) and (\ref{Modified_FRW02_SH=SBH}) are different from the formulation of a $\Lambda(t)$ model.
Note that when $S_{H} = S_{\rm{BH}}$, Eq.\ (\ref{Modified_eq_1}) reduces to $\dot{H}  =  - \frac{3}{2}  (1+ w)  H^{2}$, which is equivalent to Eq.\ (\ref{eq:Back_f}) for $f_{\Lambda} (t)=0$ in a $\Lambda(t)$ model, corresponding to standard cosmology.

\subsubsection{$T_{H}=T_{\rm{GH}}$ (Gibbons--Hawking temperature)}
\label{TH=TGH (Gibbons--Hawking temperature)}

When $T_{H}=T_{\rm{GH}}$, Eqs.\ (\ref{Modified_FRW01_equiv_2}) and (\ref{Modified_FRW02_2}) reduce to
\begin{align}
   H^{2}   &=  \frac{8 \pi G}{3} \rho  + H^{2}  \left (  1 - \frac{ S_{H} }{ S_{\rm{BH}} }  \right )      ,
\label{Modified_FRW01_TH=TGH}
\end{align}
and
\begin{align}
  \frac{ \ddot{a} }{ a }         &=   -  \frac{ 4 \pi G }{ 3}  (1+3w)  \rho   + H^{2}  \left (  1 - \frac{ S_{H} }{ S_{\rm{BH}} }  \right )     .
\label{Modified_FRW02_TH=TGH}
\end{align}
Equations\ (\ref{Modified_FRW01_TH=TGH}) and (\ref{Modified_FRW02_TH=TGH}) indicate that when $T_{H}=T_{\rm{GH}}$, the present model is equivalent to the formulation of a $\Lambda(t)$ model.
Here the extra driving term $f_{\Lambda} (t)$ is given by $H^{2}(1- S_{H}/S_{\rm{BH}})$.
In the present model, the extra driving term for the Friedmann equation is naturally equivalent to that for the acceleration equation.
Consequently, when $T_{H}=T_{\rm{GH}}$, both the background evolution of the universe and the density perturbations agree with those for the $\Lambda(t)$ model, although the theoretical backgrounds of the two models are different.
In general, a $\Lambda(t)$ model similar to $\Lambda$CDM models is favored \cite{Koma16} and, therefore, the present model for $T_{H}=T_{\rm{GH}}$ is also favored under similar conditions.

In this section, we have formulated a cosmological model based on the holographic scenario, where a general entropy and general temperature are considered.
A particular case of the present model is discussed in the next section, applying a power-law corrected entropy.

\section{Present model with a power-law corrected entropy $S_{pl}$} 
\label{The present model with a power-law corrected entropy}

In the previous section, we formulated a cosmological model based on the holographic scenario, where a general entropy and general temperature are considered,
with a consequence of the model being that the background evolution of the universe depends on the entropy but not on the temperature.
Accordingly, in this section, we introduce a particular entropy and apply it to the present model as a particular case.
(The temperature is discussed in Sec.\ \ref{Model with the Gibbons--Hawking temperature}.)

Various black hole entropies have in fact been proposed, such as a power-law corrected entropy \cite{Das2008,Radicella2010}, logarithmic corrections from loop quantum gravity \cite{LQG2004_123}, Tsallis--Cirto entropy \cite{Tsallis2012}, Tsallis--R\'{e}nyi entropy \cite{Czinner1Czinner2}, Barrow entropy \cite{Barrow2020}, and a generalized six-parameters entropy \cite{Nojiri2022}. 
These entropies are considered to be extended versions of the Bekenstein--Hawking entropy.
In this section, as an interesting example, we use the power-law corrected entropy $S_{pl}$, because this entropy gives an extra driving term corresponding to a power-law term.

The power-law corrected entropy is written as \cite{Das2008,Radicella2010}
\begin{equation}
  S_{pl}  = S_{\rm{BH}}  \left [ 1-  \Psi_{\alpha} \left ( \frac{H_{0}}{H} \right )^{2- \alpha}  \right ]    , 
\label{eq:SHpl}      
\end{equation}
where $H_{0}$ represents the Hubble parameter at the present time and $\alpha$ and $\Psi_{\alpha}$ are dimensionless constants whose values are real numbers.
In previous works \cite{Koma14,Koma15}, $\alpha$ and $\Psi_{\alpha}$ were considered to be independent free parameters, that is, $\Psi_{\alpha}$ is a kind of density parameter for the effective dark energy.
(The power-law corrected entropy is based on the entanglement of quantum fields between inside and outside the horizon \cite{Das2008}. 
The formula is summarized in Ref.\ \cite{Radicella2010}.)

We now apply the power-law corrected entropy $S_{pl}$ to the present model.
Setting $S_{H} = S_{pl}$ and substituting Eq.\ (\ref{eq:SHpl}) into Eq.\ (\ref{Modified_eq_1}) yields the simple equation 
\begin{align}
    \dot{H}  &=  - \frac{3}{2}  (1+ w)  H^{2} \left ( \frac{ S_{H} }{ S_{\rm{BH}} }  \right )    \notag \\
                &=   - \frac{3}{2}  (1+ w)  H^{2} \left [ 1-  \Psi_{\alpha} \left ( \frac{H_{0}}{H} \right )^{2- \alpha}  \right ]    \notag \\
                &=   - \frac{3}{2}  (1+ w)  H^{2} \left [ 1-  \frac{ \Psi_{\alpha} H_{0}^{2} \left ( \frac{H}{H_{0}} \right )^{\alpha} }{H^{2}} \right ]    \notag \\
                &=   - \frac{3}{2}  (1+ w)  H^{2}  + \frac{3}{2}  (1+ w)  \Psi_{\alpha} H_{0}^{2} \left ( \frac{H}{H_{0}} \right )^{\alpha} . 
\label{Modified_eq_Spl}
\end{align}
The above equation is derived from the present model with $S_{pl}$.
In fact, this equation is equivalent to an equation for a $\Lambda(t)$ model with a power-law term examined in previous works \cite{Koma11,Koma12,Koma14,Koma19}.
We can confirm that substituting a power-law term $f_{\Lambda}(t)   =   \Psi_{\alpha} H_{0}^{2}  (  H / H_{0} )^{\alpha}$ into Eq.\ (\ref{eq:Back_f}) gives Eq.\ (\ref{Modified_eq_Spl}).
The power-law term $f_{\Lambda}(t)$ for the $\Lambda(t)$ model has been examined in those previous works.
(A similar power series of $H$ for the $\Lambda(t)$ model was examined in, e.g., Ref.\ \cite{Valent2015Sola2019}.
A power-law term for other models such as CCDM models was examined in, e.g., Ref.\ \cite{Freaza2002Cardenas2020}.)

Of course, the Friedmann and acceleration equations for the present model with $S_{pl}$ are generally different from those for the $\Lambda(t)$ model with the power-law term, because of the normalized temperature, except for a special case.
(The special case, namely $T_{H} =T_{\rm{GH}}$, is discussed in Sec.\ \ref{Model with the Gibbons--Hawking temperature}.)
However, the background evolution of the universe described by Eq.\ (\ref{Modified_eq_Spl}) is always equivalent to that for the $\Lambda(t)$ model with the power-law term.
Therefore, we use solutions examined in Refs.\ \cite{Koma11,Koma14}, to discuss the background evolution of the universe.

The solution of Eq.\ (\ref{Modified_eq_Spl}) for $\alpha \neq 2$ is written as \cite{Koma14}
\begin{equation}  
    \left ( \frac{H}{H_{0}} \right )^{2-\alpha}  =   (1- \Psi_{\alpha})   \left ( \frac{a}{a_{0}} \right )^{ - \frac{3 (1+w) (2-\alpha)}{2}  }  + \Psi_{\alpha}      ,
\label{eq:Sol_HH0_power_00}
\end{equation}
where $a_{0}$ is the scale factor at the present time.
This equation reduces to the solution for $\alpha = 2$ when $\alpha \rightarrow 2$ is applied to Eq.\ (\ref{eq:Sol_HH0_power_00}).
Also, $\Lambda$CDM models in a flat FRW universe are obtained from Eq.\ (\ref{eq:Sol_HH0_power_00}), neglecting the influence of radiation.
Substituting $\alpha =0$ and $w =0$ into Eq.\ (\ref{eq:Sol_HH0_power_00}) and replacing $\Psi_{\alpha}$ by $\Omega_{\Lambda}$ yields \cite{Koma14}
\begin{equation}
 \left (  \frac{H}{H_{0}} \right )^{2}  =   (1- \Omega_{\Lambda} )   \left ( \frac{a}{a_{0}} \right )^{ - 3}  + \Omega_{\Lambda}    ,
\label{eq:Sol_H_LCDM}
\end{equation}
where $\Omega_{\Lambda}$ is the density parameter for $\Lambda$ and is given by $\Lambda /( 3 H_{0}^{2} ) $. 
In this way, the background evolution of the universe for the present model with $S_{pl}$ is equivalent to that for the $\Lambda(t)$ model with the power-law term.

In the next subsection, a temperature on the horizon is discussed, using the present model with the power-law corrected entropy $S_{pl}$.

\subsection{Present model with $S_{pl}$ for $T_{H} =T_{\rm{GH}}$} 
\label{Model with the Gibbons--Hawking temperature}

When $T_{H}=T_{\rm{GH}}$, the cosmological equations for the present model are equivalent to those for a $\Lambda(t)$ model, as shown in Eqs.\ (\ref{Modified_FRW01_TH=TGH}) and (\ref{Modified_FRW02_TH=TGH}).
Accordingly, in this subsection, we consider the present model with the power-law corrected entropy $S_{pl}$ for $T_{H} =T_{\rm{GH}}$.

Substituting Eq.\ (\ref{eq:SHpl}) into Eq.\ (\ref{Modified_FRW01_TH=TGH}) yields the Friedmann equation, written as 
\begin{equation}
 H^2      =  \frac{ 8\pi G }{ 3 } \rho    + \Psi_{\alpha} H_{0}^{2} \left ( \frac{H}{H_{0}} \right )^{\alpha}            ,                                              
\label{FRW01__Spl-TGH} 
\end{equation} 
where $S_{H} = S_{pl}$ is used. 
A calculation of $H^{2}(S_{H}/S_{\rm{BH}})$ has been performed in Eq.\ (\ref{Modified_eq_Spl}).
The second term on the right-hand side of Eq.\ (\ref{FRW01__Spl-TGH}) is an extra driving term.
Similarly, substituting Eq.\ (\ref{eq:SHpl}) into Eq.\ (\ref{Modified_FRW02_TH=TGH}) yields the acceleration equation, written as 
\begin{align}
  \frac{ \ddot{a} }{ a }     &= -  \frac{ 4\pi G }{ 3 }  ( 1+  3w ) \rho    +  \Psi_{\alpha} H_{0}^{2} \left ( \frac{H}{H_{0}} \right )^{\alpha}     ,
\label{FRW02_Spl-TGH}
\end{align}
where $S_{H} = S_{pl}$ is used. 
The second term on the right-hand side of Eq.\ (\ref{FRW02_Spl-TGH}) is an extra driving term.

We can confirm that the driving term for the Friedmann equation is equivalent to that for the acceleration equation.
Consequently, the Friedmann and acceleration equations, namely Eqs.\ (\ref{FRW01__Spl-TGH}) and (\ref{FRW02_Spl-TGH}), are equivalent to those for a $\Lambda(t)$ model with a power-law term, which is given by $f_{\Lambda}(t)   =   \Psi_{\alpha} H_{0}^{2}  (  H / H_{0} )^{\alpha}$ \cite{Koma11,Koma12,Koma14,Koma16,Koma19}.
The properties of the $\Lambda(t)$ model with the power-law term have been examined in those works.
For example, the background evolution of the universe and density perturbations have been discussed in a $(\Psi_{\alpha}, \alpha)$ plane \cite{Koma16}.
Those previous results can be applied to the present model with the power-law corrected entropy $S_{pl}$, although the theoretical backgrounds are different.

\subsubsection{Evolution of the Kodama--Hayward temperature $T_{\rm{KH}}$} 
\label{A universe at constant dynamical temperature}

Horizons of universes (including our Universe) are generally considered to be dynamic, unlike for de Sitter universes \cite{Koma19}.
In this sense, a dynamical temperature should be appropriate for discussions of thermodynamics on the dynamic horizon.
As a matter of fact, based on the present model, we can discuss thermodynamics on the dynamic horizon, e.g., by observing a dynamical temperature (different from the selected $T_{H} =T_{\rm{GH}}$).
To this end, we introduce a dynamical Kodama--Hayward temperature $T_{\rm{KH}}$, according to a previous work \cite{Koma19}.
Note that the Gibbons--Hawking temperature $T_{\rm{GH}}$ is considered to be a physical temperature used for the present model although interesting evolutions of $T_{\rm{KH}}$ are observed here.

The Kodama--Hayward temperature on the cosmological horizon of an FRW universe has been proposed \cite{Dynamical-T-20072014}, based on the works of Hayward \textit{et al}. \cite{Dynamical-T-1998,Dynamical-T-2008}.
The Kodama--Hayward temperature $T_{\rm{KH}}$ for a flat FRW universe can be written as \cite{Tu2018,Tu2019}
\begin{equation}
 T_{\rm{KH}} = \frac{ \hbar H}{   2 \pi  k_{B}  }  \left ( 1 + \frac{ \dot{H} }{ 2 H^{2} }\right )  .
\label{eq:T_KH}
\end{equation}
Here $1 + \frac{ \dot{H} }{ 2 H^{2} } > 0$ is assumed for a positive temperature in an expanding universe.
For de Sitter universes, $T_{\rm{KH}}$ reduces to $T_{\rm{GH}}$ because of $\dot{H}=0$ and, therefore, $T_{\rm{KH}}$ is interpreted as an extended version of $T_{\rm{GH}}$.
The Kodama--Hayward temperature has been recently examined, using a $\Lambda(t)$ model with a power-law term \cite{Koma19} (which is equivalent to the present model with $S_{pl}$ for $T_{H} =T_{\rm{GH}}$).
As examined in Ref.\ \cite{Koma19}, the Kodama--Hayward temperature $T_{\rm{KH}}$ is constant when the following equation is satisfied:
\begin{align}
    \dot{H} &= - 2 H^{2}  +  2 \psi H_{0}  H  ,
\label{Cosmo_T_H_mod_cst1}
\end{align}
where $\psi$ represents a dimensionless constant. 
We can confirm that substituting Eq.\ (\ref{Cosmo_T_H_mod_cst1}) into Eq.\ (\ref{eq:T_KH}) gives a constant temperature given by $T_{\rm{KH}} = \hbar \psi H_{0} /(2 \pi k_{B})$.
This universe at constant dynamical temperature should contribute to the study of horizon thermodynamics in modified FRW cosmologies, because systems at constant temperature play important roles in thermodynamics and statistical physics \cite{Koma19}.

Therefore, we examine such a universe at constant dynamical temperature.
In fact, Eq.\ (\ref{Modified_eq_Spl}) is equivalent to Eq.\ (\ref{Cosmo_T_H_mod_cst1}), under a specific condition: $\alpha=1$ and $w=1/3$ \cite{Koma19}.
We can confirm that substituting $\alpha=1$ and $w=1/3$ into Eq.\ (\ref{Modified_eq_Spl}) reduces to Eq.\ (\ref{Cosmo_T_H_mod_cst1}), where $\Psi_{\alpha}= \psi$ is also considered.
That is, the universe described by Eq.\ (\ref{Modified_eq_Spl}) satisfies a constant dynamical temperature when both $\alpha=1$ and $w=1/3$.
(The model considered here is one viable scenario, in that other models \cite{Koma19} can also satisfy Eq.\ (\ref{Cosmo_T_H_mod_cst1}).)

\begin{figure} [t] 
\begin{minipage}{0.495\textwidth}
\begin{center}
\scalebox{0.33}{\includegraphics{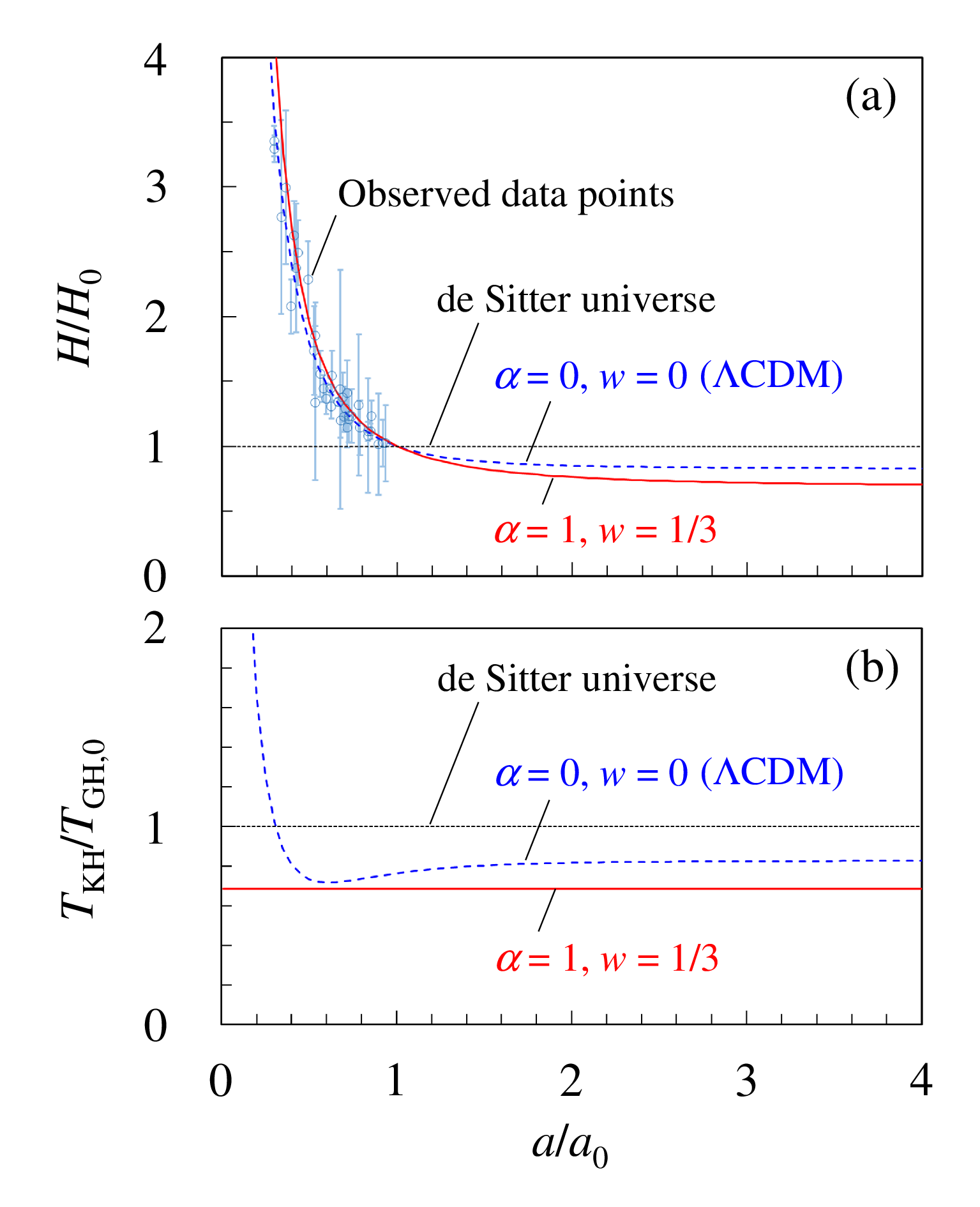}}
\end{center}
\end{minipage}
\caption{Evolution of the universe for the present model with $S_{pl}$ for $T_{H} =T_{\rm{GH}}$ for $\Psi_{\alpha}= 0.685$.
(a) Normalized Hubble parameter $H/H_{0}$.
(b) Normalized Kodama--Hayward temperature $T_{\rm{KH}}/T_{\rm{GH},0}$.
In (a), the open circles with error bars are observed data points taken from Ref.\ \cite{Hubble2017}. 
To normalize the data points, $H_{0}$ is set to $67.4$ km/s/Mpc from Ref.\ \cite{Planck2018}, as examined in Refs.\ \cite{Koma14,Koma15,Koma16,Koma17,Koma19}.
In (b), $T_{\rm{KH}}$ is normalized by $T_{\rm{GH},0}$, namely the Gibbons--Hawking temperature at the present time \cite{Koma19}.
The plots for [$\alpha =0$, $w=0$] and [$\alpha =1$, $w=1/3$] correspond to the $\Lambda$CDM model and a universe at constant dynamical temperature, respectively.
See the text.
 }
\label{Fig-H-a}
\end{figure}

We now observe the evolution of the normalized Hubble parameter and the normalized Kodama--Hayward temperature.
Similar forms of evolution have been examined in Ref.\ \cite{Koma19}.
In Fig.\ \ref{Fig-H-a}, to examine typical results, $\Psi_{\alpha}$ is set to $0.685$, which is equivalent to $\Omega_{\Lambda}$ for the $\Lambda$CDM model from the Planck 2018 results \cite{Planck2018}.
Therefore, the plots for [$\alpha =0$, $w=0$] are equivalent to those for the $\Lambda$CDM model.
The plots for [$\alpha =1$, $w=1/3$] correspond to a universe at constant dynamical temperature.
The evolution of the normalized Hubble parameter is obtained from Eq.\ (\ref{eq:Sol_HH0_power_00}).
Also, using Eq.\ (\ref{eq:T_KH}), the normalized Kodama--Hayward temperature can be written as \cite{Koma19}
\begin{equation}
\frac{T_{\rm{KH}}}{ T_{\rm{GH},0} } =  \frac{H}{H_{0}}  \left ( 1 + \frac{ \dot{H} }{ 2 H^{2} }\right )   ,
\label{eq:TKH_TGH0}      
\end{equation}
where $T_{\rm{GH},0}$ represents the Gibbons--Hawking temperature at the present time, given by $T_{\rm{GH},0} =  \hbar H_{0} / (2 \pi  k_{B})$. 
The normalized $T_{\rm{KH}}$ for the present model with $S_{pl}$ is calculated as follows.
Substituting Eq.\ (\ref{Modified_eq_Spl}) into Eq.\ (\ref{eq:TKH_TGH0}) and substituting Eq.\ (\ref{eq:Sol_HH0_power_00}) into the resultant equation and performing several calculations yields \cite{Koma19}
\begin{align}
\frac{T_{\rm{KH}}}{ T_{\rm{GH},0} }    &=  \frac{(1-3w) (1- \Psi_{\alpha})  (a/a_{0})^{ - \gamma}    + 4 \Psi_{\alpha}   }{4  \left [ (1- \Psi_{\alpha})  (a/a_{0})^{ - \gamma}  + \Psi_{\alpha}  \right ]^{\frac{1-\alpha}{2-\alpha}} } ,    
\label{TKH_TGH0_power}
\end{align}
where $\gamma$ represents $3 (1+w) (2-\alpha) /2$ and $\alpha \neq 2$ is considered.
For details of the calculations, see Ref.\ \cite{Koma19}.
When both $\alpha =1$ and $w=1/3$, this equation reduces to a constant value given by $T_{\rm{KH}}/T_{\rm{GH},0}=\Psi_{\alpha}$.
In this figure, $H/H_{0} =1$ and $T_{\rm{KH}}/T_{\rm{GH},0} =1$ for a de Sitter universe are also plotted, where $H$ is set to $H_{0}$ for simplicity.

As shown in Fig.\ \ref{Fig-H-a}, the normalized $H$ for [$\alpha=1$, $w=1/3$] and [$\alpha=0$, $w=0$] decreases with $a/a_{0}$, unlike for the de Sitter universe.
However, the normalized $T_{\rm{KH}}$ for [$\alpha=1$, $w=1/3$] is constant during the evolution of the universe, as for the de Sitter universe.
In this way, we can confirm that the dynamical temperature for [$\alpha=1$, $w=1/3$] is constant, although the Hubble parameter varies with time as for the $\Lambda$CDM model.

Of course, our Universe should be different from a universe at constant dynamical temperature.
However, this universe is expected to be a good model for studying relaxation processes of the universe and horizon thermodynamics \cite{Koma19}.
Based on the present model, we can examine not only the evolution of the universe but also the thermodynamics on dynamic horizons of modified FRW universes.
By extending this model, we may examine the relationship between holographic entanglement entropy \cite{RyuTakayanagi2006,deSitter23,Takayanagi2023} and thermodynamic entropy on the dynamic horizon \cite{Koma19}.

In this section, we have only discussed a particular case of the present model.
Various forms of the entropy and temperature should also be applied to the present model.
Those tasks are left for future research.

\section{Conclusions}
\label{Conclusions}

We phenomenologically formulated a cosmological model based on holographic scenarios in a flat FRW universe, whose horizon is assumed to have a general entropy $S_{H}$ and general temperature $T_{H}$.
To formulate the model, we also assumed that a holographic-like connection and Padmanabhan's holographic equipartition law can be applied to $S_{H}$ and $T_{H}$.
Based on these assumptions, we derived the Friedmann and acceleration equations from the holographic-like connection and holographic equipartition law, respectively.
The derived Friedmann and acceleration equations are slightly complicated because these two equations include both the normalized $S_{H}$ and normalized $T_{H}$.

Using the Friedmann and acceleration equations, we formulated a cosmological model based on the holographic scenario.
It is found that these two equations lead to a simple equation, corresponding to a similar equation that describes the background evolution of the universe in time-varying $\Lambda (t)$ cosmologies.
The simple equation does not depend on $T_{H}$, because $T_{H}$ included in the two equations cancel each other.
That is, the background evolution of the universe for the present model depends on the selection of $S_{H}$ but does not depend on the selection of $T_{H}$.
These results imply that the holographic-like connection should be consistent with Padmanabhan's holographic equipartition law through the present model and that the entropy plays a more important role.
When the Gibbons--Hawking temperature $T_{\rm{GH}}$ is selected as $T_{H}$, the cosmological equations (namely, the Friedmann and acceleration equations) are found to be equivalent to those for a $\Lambda(t)$ model, although the theoretical backgrounds are different.

Finally, we examined a particular case of the present model, applying a power-law corrected entropy.
The background evolution of the universe agrees with that for a $\Lambda(t)$ model with a power-law term.
(When $T_{H}=T_{\rm{GH}}$, the cosmological equations for these two models are the same and, therefore, density perturbations are also the same.)
Under a specific condition, the present model with the power-law corrected entropy can describe a universe at constant dynamical temperature.

The holographic-like connection, Padmanabhan's holographic equipartition law, and several assumptions used here have not yet been established.
Therefore, detailed studies are needed.
However, those scenarios are likely consistent with each other through the present model, as if they are one scenario.
Based on the present model, we should be able to examine the evolution of modified FRW universes and the thermodynamics on dynamic horizons in holographic cosmology.

\appendix

\section{Continuity equation for the present model}
\label{Continuity equation}

In this appendix, a modified continuity equation for the present model is examined.
For this, we introduce a general formulation for cosmological equations, according to previous works \cite{Koma456,Koma16}. 
The general Friedmann, and acceleration, and continuity equations for a flat FRW universe can be written as
\begin{align}
     H^{2}  &=  \frac{ 8\pi G }{ 3 } \rho  + f(t)   ,
\label{eq:A_mFRW01}
\end{align}
\begin{align}
\frac{ \ddot{a} }{ a }  &=  -  \frac{ 4\pi G }{ 3 } ( 1 +  3w) \rho   + g(t)   , 
\label{eq:A_mFRW02}
\end{align}
\begin{equation}
       \dot{\rho} + 3  H  ( 1  +  w )  \rho   =  \frac{3}{4 \pi G} H \left(  - f(t) -  \frac{\dot{f}(t) }{2 H }  +  g(t)      \right )     ,
\label{eq:A_drho}
\end{equation}
where $f(t)$ and $g(t)$ represent extra driving terms for the Friedmann and acceleration equations, respectively \cite{Koma456,Koma16}. 
The continuity equation can be derived from the Friedmann and acceleration equations because two of the three equations are independent.
When $f(t) = g(t) = f_{\Lambda}(t)$, the three equations reduce to cosmological equations for a $\Lambda(t)$ model. 
Accordingly, the general Friedmann and acceleration equations reduce to Eqs. (\ref{eq:General_FRW01_f_0}) and (\ref{eq:General_FRW02_f_0}), respectively.
Also, the continuity equation for the $\Lambda(t)$ model is given by 

\begin{equation}
      \dot{\rho} + 3  H  ( 1  +  w )  \rho   =  - \frac{3}{8 \pi G}  \dot{f}_{\Lambda} (t)   .
\label{eq:drho_L(t)}
\end{equation}
This equation has a non-zero right-hand side, as for Eq.\ (\ref{eq:A_drho}).
The non-zero right-hand side is discussed later.

We now examine the continuity equation for the present model. 
From Eqs.\ (\ref{Modified_FRW01_equiv_2}) and (\ref{Modified_FRW02_2}), the Friedmann and acceleration equations are written as   
\begin{align}
  H^{2}     &=  \frac{8 \pi G}{3} \rho   + H^{2}   \left [ 1-  \left ( \frac{ S_{H} }{ S_{\rm{BH}} }  \right )   \left ( \frac{T_{H}}{T_{\rm{GH}}} \right )  \right ]  ,
\label{Modified_FRW01_equiv_3}
\end{align}
\begin{align}
  \frac{ \ddot{a} }{ a }            &=   -  \frac{ 4 \pi G }{ 3}  (1+3w)  \rho \left (  \frac{T_{\rm{GH}}}{T_{H}} \right )      + H^{2}  \left (  1 - \frac{ S_{H} }{ S_{\rm{BH}} }  \right )     .
\label{Modified_FRW02_3}
\end{align}
For simplicity, the extra driving terms for the Friedmann and acceleration equations are replaced by $\tilde{f}(t)$ and $\tilde{g}(t)$, respectively, which are given as
\begin{align}
  \tilde{f}(t)  &=   H^{2}   \left [ 1-  \left ( \frac{ S_{H} }{ S_{\rm{BH}} }  \right )   \left ( \frac{T_{H}}{T_{\rm{GH}}} \right )  \right ]  ,
\label{f(t)_Present}
\end{align}
\begin{align}
  \tilde{g}(t)  &=  H^{2}  \left (  1 - \frac{ S_{H} }{ S_{\rm{BH}} }  \right )     .
\label{g(t)_Present}
\end{align}
Using Eqs.\ (\ref{Modified_FRW01_equiv_3}) and (\ref{Modified_FRW02_3}), applying $H=\dot{a}/a$, and performing several calculations yields
\begin{equation}
       \dot{\rho} + 3  H  ( 1  +  \tilde{w} )  \rho   =  \frac{3}{4 \pi G} H \left(  - \tilde{f}(t) -  \frac{\dot{\tilde{f}}(t) }{2 H }  +  \tilde{g}(t)      \right )     , 
\label{eq:drho_Present}
\end{equation}
where $\tilde{w}$ represents a modified equation-of-state parameter, which is given by
\begin{equation}
       \tilde{w} =  w - \frac{ ( 1 + 3 w )   \left ( 1-  \frac{ T_{\rm{GH}}}{T_{H}} \right )  }{3}  .
\label{eq:w_mod}
\end{equation}
Equation\ (\ref{eq:drho_Present}) is the continuity equation for the present model.
This equation is equivalent to the formulation of Eq.\ (\ref{eq:A_drho}) and has a similar non-zero right-hand side.
Note that $\tilde{w}$ is a modified equation-of-state parameter and, when $T_{H} = T_{\rm{GH}}$, $\tilde{w}$ reduces to $w$.

In this way, the continuity equation for the present model has a non-zero right-hand side.
A similar non-zero term is included in the continuity equation for other cosmological models, such as bulk viscous cosmology \cite{Weinberg0,BarrowLima,BrevikNojiri,EPJC2022} and energy exchange cosmology \cite{Barrow22,Wang0102,Dynamical20052013}, as discussed in Ref.\ \cite{Koma456}.
In bulk viscous cosmology, an effective continuity (conservation) equation can be obtained from an effective description of the equation of state, using a single fluid. 
The effective continuity (conservation) equation can be written as \cite{Koma456} 
\begin{equation}
  \dot{\rho} + 3 H ( 1 + w_{e} ) \rho =0  ,
\label{eq:drho_bulk}
\end{equation}
where $w_{e}$ is an effective equation-of-state parameter which includes the non-zero right-hand term of the original continuity equation.
In contrast, energy exchange cosmology assumes the transfer of energy between two fluids \cite{Barrow22}, e.g., the interaction between dark matter and dark energy \cite{Wang0102} and the interaction between matter and dynamical vacuum energy \cite{Dynamical20052013}.
(Energy exchange cosmology is equivalent to the formulation of a $\Lambda(t)$ model.)
In energy exchange cosmology, the continuity equation for each fluid has a similar non-zero term on the right-hand side.
For example, using a dynamical vacuum (dark) energy term $\Lambda(t)/3$, the continuity equations for matter `$m$' and vacuum (dark) energy `$\Lambda$' can be written as \cite{Dynamical20052013}
\begin{equation}
\dot{\rho}_{m}            + 3 H ( 1 + w_{m}) \rho_{m}                       = - - \frac{1}{8 \pi G}   \dot{\Lambda}   ,
\label{eq:drho_L(t)_matter}
\end{equation}
\begin{equation}
\dot{\rho}_{\Lambda} + 3 H ( 1 + w_{\Lambda }) \rho_{\Lambda}  =  \frac{1}{8 \pi G}  \dot{\Lambda}   .
\label{eq:drho_L(t)_vacuum}
\end{equation}
The two non-zero right-hand sides are totally cancelled because the total energy of the two fluids is conserved.
In this study, $f_{\Lambda} (t)$ for a $\Lambda(t)$ model corresponds to $\Lambda(t)/3$.

As examined above, when $T_{H}=T_{\rm{GH}}$ is considered, the present model is equivalent to the formulation of a $\Lambda(t)$ model.
Therefore, the continuity equation, namely Eq.\ (\ref{eq:drho_Present}), reduces to Eq.\ (\ref{eq:drho_L(t)}), where $f_{\Lambda} (t)$ is given by $H^{2}(1- S_{H}/S_{\rm{BH}})$.
In this case, the non-zero right-hand side of the continuity equation should imply the interaction between matter and dynamical vacuum (dark) energy.
Alternatively, the non-zero right-hand side may be interpreted as the interchange of energy between a cosmological horizon and the bulk.
However, when $T_{H} \neq T_{\rm{GH}}$, not only such the interactions but also an effective description used for bulk viscous cosmology should be considered.
These results may imply that the present model for $T_{H}=T_{\rm{GH}}$ is favored from a viewpoint of simplicity.
Further studies are needed and those tasks are left for future research.

\end{document}